**Web3 and the State: The Indian state's redescription of blockchain**

Debarun Sarkar*, Independent Researcher, debarun@outlook.com

Cheshta Arora*, Western Norway Research Institute, car@vestforsk.no

**Direct comments to** Cheshta Arora, car@vestforsk.no

**Abstract:** The article closely reads a discussion paper by the National Institution for Transforming India (NITI) Aayog and a strategy paper by the Ministry of Electronics and Information Technology (MeitY) advocating non-financial use cases of blockchain in India. By noting the discursive shift from transparency to trust to adjustably transparent enacted in these two documents, and consequently the Indian state's redescription of blockchain, the paper foregrounds how blockchain systems are being designated as "decentral" but have recentralizing effects where the state reinvents and re-establishes itself as an intermediary. The paper illustrates how discursive shifts concerning trust, transparency, (de)centralization and (dis)intermediation are crucial sites for investigating redescriptions of emerging sociotechnical systems.

*Both authors have contributed equally to the research and development of this text.

## I. Introduction:

The way in which blockchain has captured the interest of the market in the last few years is being increasingly documented and critically analysed from different perspectives (Brekke, 2018; Herian, 2018; Jacobetty and Orton-Johnson, 2022). However, there is a skewed focus on mapping the interest of capitalists and entrepreneurs in the global north with only limited work on mapping techno-financial experimentation in the global south (Campbell-Verduyn and Giumelli, 2022) as well as how blockchain captures the imagination of other political actors. This skewed focus on capitalists and entrepreneurs, and blockchain's potential to reinvent finance, obfuscates how another important actor[1], the state, aims to re-describe blockchain. Field-based studies of blockchain projects are lacking, with few exceptions (Faustino, 2019; Rodima-Taylor, 2021). Discursive studies of the national policy or strategy documents are also missing. This omission is jarring, considering that until 2018, at least 100 projects were underway in 30 countries to transform government systems (Jun, 2018). At the time of writing, blockchain adoption and deployment in India ranged across a few use cases that involved using blockchain to store documents and certificates relating to birth, death, caste, land, judicial documents, and logistics.[2]

This increasing interest of states in blockchain projects is attributed to blockchain's 'consensus mechanisms' that help decide if the public data is genuine, which makes blockchain "an optimal technology for dealing with public data […] with the potential to replace social organizations" (Jun, 2018). As an advanced distributed ledger technology (DLT), it promises data reliability across remote machines (Jacobetty and Orton-Johnson, 2022). Its technical possibilities, however, do not satisfactorily explain why states are invested in experimenting with and adopting blockchain projects (Jun, 2018) and how the technology itself is redescribed.

To focus on redescription is to consider the material-semiotic (Law, 2019) nature of things, objects and bodies around us. Redescription is part of a broader constellation that forms the method of script analysis which argues that technological objects "define a framework of action together with the actors and the space in which they act" (Akrich, 1992, p. 209). A script analysis studies inscription (what is written into an object?), description (what is occurring between objects and bodies in a setting?), prescription (what is allowed?) and subscription (how is it read?) vis-à-vis a technological object (Fallan, 2008). By mapping how blockchain is being redescribed (or reinscribed), we make a distinction between the inscription enacted by Nakamoto (2008) in the context of Bitcoin and the Indian state's redescription (or re-inscription) of that initial moment of inscription by the inventor. While re-inscription or redescription are conceptually interchangeable, we use redescription to hint at the reshuffling or reordering being enacted and the respective world being engendered therein (Faustino, 2019).

---

[1] It should be noted that 'actor' is a shorthand for an assemblage of human and nonhuman processes that at times coagulate and attain a distinct identity. In this way, the state is both an actor and a network.
[2] See https://blockchain.gov.in/

A blockchain network as a socio-technical assemblage of human and non-human components (Faustino, 2019) is an emerging site of political contestation and possibility (Gloerich *et al.*, 2018, p. 8). At the same time, the anthropological view of the state foregrounds the state as a contested institutional form that comes into being through heterogeneous ideological and material aspects of everyday social practices, relations, actors, objects, technologies, and processes (Sharma and Gupta, 2006, p. 8). From within this contested and contingent nature of both blockchain as well as the state, it is worthwhile to unpack how the promise of blockchain is reworked when it is envisioned as the state's infrastructure, as well as how the state reinvents itself in light of the technology's alleged claims of transparency, disintermediation, decentralization and distribution.

The critical history of the Internet is ripe with political imaginaries of de/recentralization. These imaginaries range from "abstract claims about inherent political properties of internet protocols" to "social factors, institutional procedures and material aspects of internet infrastructure" (Beuster *et al.*, 2022). Bodó et al. (2021) make a useful distinction between three characteristics of decentralization — 1) as a principle for design, 2) an end goal to have decentralizing effects, and 3) a claim that designates a system as decentral. Based on this distinction, we can differentiate between blockchain experiments that are intended to be decentral in both their design and their effect (Husain *et al.*, 2020) versus experiments that are designated as decentral but have a recentralizing socio-technical design or effects.

In this article, we focus on tracing discursively the third aspect of decentralization and interrogate the ways in which a system is designated as decentralised but has recentralizing effects.

We offer a critical reading of two documents — *Blockchain: The India Strategy, Part I* published by the National Institution for Transforming India (NITI) Aayog in January 2020 and *National Strategy on Blockchain* published by the Ministry of Electronics and Information Technology (MeitY), Government of India in January 2021. With these two texts, a constellation of other texts also appeared, *Tamil Nadu Blockchain Policy 2020* published by the Information Technology Department of the Government of Tamil Nadu, the draft of the Union government bill, *Banning of Cryptocurrency & Regulation of Official Digital Currency Bill, 2019* (Government of India, 2019) which has gone through revisions and backtracks over the years, and the Supreme Court judgement in *Internet and Mobile Association of India v. Reserve Bank of India* (2018). These other ancillary texts provide a sense of the emerging genealogy of the state's encounter with this new technological development, and its efforts to comprehend, govern and domesticate it. However, we only focus on mapping the discursive shifts being enacted in the two documents by NITI Ayog and MeitY in this article.

Methodologically, we borrow from critical policy studies literature, that investigates state-originated documents as an "ideologically constructed product of political forces [which] in spite of itself, embodies incoherences, distortions, structured omissions and negations" (Codd, 1988).

A critical analysis of state documents can foreground the ways in which "infrastructural control can serve as proxies to regain (or gain) control or manipulate the flow of money, information, and the marketplace of ideas in the digital sphere" (Musiani *et al.*, 2016, p. 4). While states are being called upon to arrive at an overarching regulatory framework concerning blockchain and cryptocurrencies (Bains *et al.*, 2022; Sarkar, 2022; Taylor, 2018), the paper makes opening moves to think through the foreclosures in state documents on blockchain.

The paper argues that blockchain is being re-described by the Indian state to further disintermediation in the workings of the state—where the state gets rid of its own heterogenous organs to centralize power. This rests on a discursive shift away from transparency to trust to blockchain *as adjustably transparent* — as is laid out in the analysis of the two texts in section 3. However, this political imagination of disintermediation falls short of rethinking new relationalities between the state and citizens that blockchain's promise of decentralization and transparency could enable.[3]

In the second section, we briefly trace the history of the Indian state's ambivalent response to blockchain and its attempt to disassociate blockchain from cryptocurrency. The following core section of the paper does a close reading of two texts to map the discursive shift from transparency to trust to "adjustably transparent" which is at the core of the Indian state's re-description of Blockchain.

## II. The Indian state's response to cryptocurrency and blockchain

Cryptocurrency emerged in response to the behaviour of centralized financial authorities that led to the 2008 crisis and "with the monopoly of central banks crumbling, the very definition of money [was] up for grabs" (Gloerich *et al.*, 2018): what money is, how it is valued, who regulates it and assures its promissory value (Nakamoto, 2008).

This explicit financial use-case of blockchain technology was soon supplanted by more abstract notions of value introduced by the development of smart contracts in second-generation projects (Vigliotti, 2021). Smart contracts led to the possibility of executing what is referred to as decentralized finance, and other forms of complex derivative transactions, in a decentralized manner (Chen and Bellavitis, 2020).

This departure marked a moment in the development of blockchain technology wherein the ledger became more than a mere record of transactions (Savelyev, 2016; Hassan and De Filippi, 2017; De Filippi and Hassan, 2016). With second-generation projects like Ethereum, the code also became proof and law, recording complex contracts such as the transfer of property, loans, mortgages, and complex derivatives and facilitating various decentralized applications from video games to social media (Wu, 2019). These technological developments made it possible to distinguish between financial use-cases of blockchain, such as cryptocurrencies, and non-financial use-cases of blockchain as distributed ledger technology (DLT).

---

[3] Technical limitations of decentralization still remain with the challenge of 51% attack for example (Kroll *et al.*, 2013).

Like several other states, the Indian state's response to cryptocurrency and non-financial use-cases of blockchain has not been the same. It is important to mark out the variety of these responses to underscore that the state's use of blockchain rests on its re-description of blockchain as DLT while enacting certain erasures concerning its financial use cases. This distinction between cryptocurrency and non-financial use cases of blockchain is reflected in the manner in which different organs of the state have responded to the developments in blockchain and cryptocurrency. We give a brief account of these developments below.

*The ban*

On April 5, 2018, the Reserve Bank of India made a press release that included a paragraph on cryptocurrencies referred to as virtual currencies (VC) and notified that "entities regulated by RBI shall not deal with or provide services to any individual or business entities dealing with or settling VCs" (Reserve Bank of India, 2018b).

The next day, on April 6, 2018, the Reserve Bank of India published a notification addressed to all banking and payments service providers for "maintaining accounts, registering, trading, settling, clearing, giving loans against virtual tokens, accepting them as collateral, opening accounts of exchanges dealing with them and transfer/receipt of money in accounts relating to purchase/sale of" (Reserve Bank of India, 2018a) cryptocurrencies. This wasn't the first time that the RBI had published a circular about cryptocurrencies as the circular noted that "Reserve Bank has, repeatedly through its public notices on December 24, 2013, February 01, 2017, and December 05, 2017, cautioned users, holders and traders of virtual currencies, including Bitcoins, regarding various risks associated in dealing with such virtual currencies" (Reserve Bank of India, 2018a).

What was different from the previous notifications, though, was that this notification essentially sanctioned a banking blockade against the circulation of cryptocurrency. It cut off the exchange of fiat and cryptocurrencies by explicitly targeting cryptocurrency trading exchanges. This does not mean that cryptocurrency was suddenly not being traded by Indian citizens or that all exchanges were shut down. Instead, the trade moved to peer-to-peer transactions facilitated by various exchanges as escrows.

*The reversal*

RBI's ban was challenged by the Internet and Mobile Association of India as well as cryptocurrency exchange platforms. It took roughly two years for the ban to be overturned by the Supreme Court of India. The Supreme Court's judgement remains an insightful document to map different descriptions of crypto and DLT — as currency, commodity, stock, property, funds, digital representation of value, digital means of payment, digital cash, medium of exchange, universal money, crypto as a source of financial, legal and security-related risks etc. An analysis of the judgement and different re-descriptions of crypto by various actors being played out therein remains beyond the scope of the paper. The verdict did not resolve these contesting descriptions of crypto. Instead, it relied on crypto exchange platforms' right to free trade enshrined in Article 19(1)(g) of the constitution. RBI's circular had restricted the exchanges' access to banking services and was thus found to be in violation of Article 19(1)(g).

One significant distinction made by the petitioners which is relevant to our discussion is the distinction between cryptocurrency and blockchain. While laying their cards on the table, the petitioner had pointed out that "it is a paradox that blockchain technology is acceptable to RBI, but cryptocurrency is not" (Supreme Court of India, 2018, p. 36). The paradox was rejected by the bench on the grounds that "there is nothing irrational about the acceptance of a technological advancement/innovation, but the rejection of a by-product of such innovation. There is nothing like a take it or leave it option" (Supreme Court of India, 2018, p. 138). While the ambiguity concerning cryptocurrencies and their legal status is retained in the judgement, blockchain as a technological advancement is acknowledged and affirmed *tout court*. Such a dislodging of cryptocurrency from blockchain is not just the erasure of the economic implications but also the political challenge that these advances herald.

The overturning of the RBI ban by the Supreme Court led to a renewed activity in the industries around cryptocurrency particularly with the re-opening of the exchanges which allowed for bank transfers between the exchanges and the banks with certain limitations. This was also followed by the increased visibility of the crypto exchanges in the mainstream media, with advertisements invading various outlets and high-profile advertisement slots.

While this unfettered enthusiasm coincided with a bull market of cryptocurrencies following the COVID-19 lockdowns, contradictory signals from different organs of the state started to be sent to media houses. Accompanying these mixed signals was the impending fear of passing the *Banning of Cryptocurrency & Regulation of Official Digital Currency Bill, 2019* (Government of India, 2019). The draft of the bill which has been circulating for the last few years argued for a direct ban but had not been tabled yet. Even the Supreme Court judgement noted the existence of the bill and the government's flip-flop on the issue as a sign of ambivalence.

However, within this atmosphere of fear and uncertainty, the banks started to cut off their payment networks and services with the cryptocurrency exchanges. This led to a notification by the RBI on May 31, 2021, which noted explicitly

> "certain banks/ regulated entities have cautioned their customers against dealing in virtual currencies by making a reference to the RBI circular DBR.No.BP.BC.104/08.13.102/2017-18 dated April 06, 2018. Such references to the above circular by banks/ regulated entities are not in order as this circular was set aside by the Hon'ble Supreme Court on March 04, 2020 in the matter of Writ Petition (Civil) No.528 of 2018 (Internet and Mobile Association of India v. Reserve Bank of India). As such, in view of the order of the Hon'ble Supreme Court, the circular is no longer valid from the date of the Supreme Court judgement, and therefore cannot be cited or quoted from." (Reserve Bank of India, 2021)

As the Supreme Court noted in its judgement, there has been a considerable amount of activity among various organs of the state in trying to comprehend and in turn consider the implications of the technological developments of cryptocurrency and blockchain. This has been accompanied by a burgeoning economy of start-ups in India ranging from cryptocurrency exchanges to organisations developing scaling solutions for Ethereum such as Polygon[4].

Publication of strategies by various organisations of the state give a sense of how these new techno-economic shifts are being re-described by different actors, and through what manoeuvres. In the next section, we present an analysis of two documents that re-describe blockchain as a decentralized managerial technology of the state with recentralizing effects.

---

[4] While the key founders of Polygon are Indian, the founders do not associate the company with any one national identity claiming that it's a "decentralized network with no headquarters", although it is officially registered in the British Virgin Islands (Mansur, 2022).

### III. Case Studies

*The NITI Aayog's Report*

The National Institution for Transforming India (NITI) Aayog (lit. 'Policy Commission') is the primary policy think-tank of the Government of India. It was founded in 2015 and replaced the Planning Commission, India's 64-year-old apex policy-making body. A comprehensive account of what this transition entailed in practice is still missing. Before its demise in 2015, the Planning Commission was central to independent India's political and economic visions and had already been at the centre of multiple critiques which we cannot discuss here. Briefly, at the time of its demise, it was already established that the commission had failed to realize the visions of the anti-colonial struggle and had displayed centralizing tendencies in its planning and functioning with an overbearing regulatory oversight.

The demands and plans to restructure the Planning Commission were already underway and the NITI Aayog emerged amidst these contestations. The transition to NITI Aayog and its political and economic implications are still contested. In its narrative, NITI Aayog sets itself against the established failures of the Planning Commission. Its roles and responsibilities have been clubbed under four major heads: 1) fostering cooperative federalism, 2) formulating strategic visions and long-term policies and programme framework for the macro-economy and different sectors, 3) acting as a knowledge and innovation hub, and 4) providing a platform for inter-departmental coordination (Rao, 2015).

However, sceptics have noted that NITI Aayog is not a shift away but a move towards further centralization. It increases control of the Union of India over the states and results in further entrenchment of neoliberal policies at the Union level by incentivizing public-private policies and increasing inter-state competition to attract private capital (Patnaik, 2015). It is important to note that NITI Aayog has no constitutional power, and its reports and suggestions hold no executive power except as recommendations. Despite this, NITI Aayog's discursive production must be taken seriously because of its peculiar configuration as a replacement for the Planning Commission (Patnaik, 2015) which makes it distinct from other publicly funded think tanks.

This quick background to NITI Ayog is relevant to map how it frames the adoption of blockchain in India in Part 1 of the discussion paper titled "Blockchain: The India Strategy – Towards Enabling Ease of Business, Ease of Living and Ease of Governance". It published the first part in January 2020 and the second part of the report was yet to be published at the time of writing and revising this paper in early 2024.

At the outset, the technology in the discussion paper is tied to the motto that has been central to the ruling party's governance approach in the last few decades: ease of collaboration for enterprise, and ease of living for citizens. It is divided into six main chapters and envisioned to be a 'pre-read' to "implementing a blockchain system in India" because, as the Foreword warns, "the technology may not be universally more efficient and thus specific use cases need to be identified where it adds value and those where it does not" (NITI Aayog, 2020, p. 5).

The discussion paper does not shy away from emphasizing that the technology "has the potential to revolutionize interactions between governments, businesses and citizens in a manner that was unfathomable just a decade ago" and "is unique in its foundation nature" with the "potential to unlock new sources of efficiency and value" (NITI Aayog, 2020, p. 6). Two main visions are identified vis-à-vis ease of business and ease of living that blockchain can help address: "'self-regulation' to ensure ease of doing business by allowing entities to interact through a *trusted medium* with a reduced dependency on cumbersome regulatory oversight and compliance" and "empowering citizens through features of *transparency, decentralization, and accountability…*" (NITI Aayog, 2020, p. 6 emphasis added).

Blockchain as a technology is explained in an Appendix at the end of the NITI Aayog discussion paper. The Introduction as well as the structure of the discussion paper show that one of the primary challenges that the authors hoped to address via the discussion paper was to re-narrate blockchain as a governance infrastructure. The Introduction acknowledges that blockchain does not lend itself as governance infrastructure easily and one of the tasks of the discussion paper is to "demystify and improve the understanding of amenability of blockchain to specific use-cases" and present the "functional view of blockchain" based on a simple decision-making framework developed in chapter 2 (NITI Aayog, 2020, p. 6) and to advocate its potential for governance.

NITI Aayog's redescription of the technology pivots upon two descriptive moves: First, it establishes blockchain as "the new trust paradigm". Second, it "removes transaction from the context of finance" and insists upon a meaning of transaction as it is "commonly defined" as "the act of carrying out or conducting a deal or exchange to a conclusion or settlement" (NITI Aayog, 2020, p. 9).

NITI Aayog's redescription of blockchain begins by evoking Friedman's (1998) Introduction to the reprint of Leonard Read's 1958 essay "I, Pencil". Friedman is used as an authoritative citational source to introduce the "invisible hand of the market" which "makes something as simple as a pencil" (NITI Aayog, 2020, p. 8) possible. From NITI Aayog's point of view, this 'invisible hand of the market', however, is riddled with disputes and disagreements that need trusted intermediaries to "ensure that the entities adhere to the commonly understood 'rules of the game'" (NITI Aayog, 2020, p. 8). Traditionally, the paper notes, governments have performed this role of a mediator and ensured contract enforcement through mechanisms of deterrence and penalty but governments' role as a mediator is also burdened with inefficiencies. Thus, enter blockchain — a technology that *represents* "an impartial and trusted facilitator", a code-based "trusted intermediary" with "the power to enforce rules" (NITI Aayog, 2020, p. 8).

In re-describing blockchain as a code-based intermediary to facilitate self-regulation, NITI Aayog does not envision that the state will recede, or outsource its function to its technological counterpart. Instead, this re-description is essential to introduce the state back into the network as a 'player' (NITI Aayog, 2020, p. 14). After describing blockchain as a trust-based system, the authors note that the Indian state has been failing in its "processes to ensure trust" (NITI Aayog, 2020, p. 10) which the use of blockchain as a trusted intermediator can help address by ensuring "less government and more governance"— which is the popular mantra underpinning the ruling government's policy approach since 2014.

It should be noted here that while blockchain is quite commonly presented as a new 'trust-based system', as the discussion paper too establishes, this description is not innocent and discursively produced which has consequences. For instance, the famous Bitcoin technical paper (Nakamoto, 2008) begins by acknowledging that the trust-based systems underpinning all commerce on the Internet are inherently weak and unreliable. For Nakamoto (2008), Bitcoin is a response to this weakness and envisions "an electronic payment system based on cryptographic proof instead of trust" (Nakamoto, 2008, p. 1).

The Bitcoin technical paper (Nakamoto, 2008) concerned itself with the problems of double spending — where the same single digital token can be spent more than once — in an electronic payments system that made it necessary to have a trusted party overseeing the transaction. To address the problem of double-spending, and to remove any dependence upon a trust-based system, it relied on introducing absolute transparency into the system where "transactions must be publicly announced" (Nakamoto, 2008, p. 2) to avoid any reliance on a trusted third party.

On the contrary, NITI Aayog's re-description of blockchain as a trust-based intermediary — rather than, say, a transparent intermediary — is at the heart of re-describing blockchain as a governance infrastructure. While in the Bitcoin technical paper, blockchain technology enabled a peer-to-peer transaction based on a transparent cryptographic proof instead of trust, in NITI Aayog's re-description, blockchain technology involves simplifying the trusted systems by reintroducing a code-based trusted intermediary in the peer-peer transaction (which otherwise has been eroded given the increasing complexity of the trust systems) (NITI Aayog, 2020, p. 9).

This is accompanied by another narrative move where the meaning of transaction is evocatively disassociated from the context of finance to its "commonly defined" notion as "the act of carrying out or conducting a deal or exchange to a conclusion or settlement… facilitated by 'trust systems' and intermediaries" (NITI Aayog, 2020, p. 9).

On page 11, after establishing the narrative move laid out above, the discussion paper finally asks, 'What is Blockchain?'. At this point in the narrative, NITI Aayog can confidently define blockchain as "a database, which is distributed, *adjustably* transparent, highly secure, and immutable" (NITI Aayog, 2020, p. 11, emphasis added).

With this definition, blockchain has been re-described in the report as a governance technology. The most interesting part of this definition is the qualifier "adjustably transparent" (as opposed to the absolute transparency underpinning the Bitcoin paper). The qualifier attests that in this vision of blockchain the records will be "made visible to *relevant stakeholders"*. Thus, re-enters the State.

The discursive addition of the qualifier adjustably transparent brings back the state as the gatekeeper. Blockchain is a code-based intermediary accessible to relevant stakeholders only via the entry/exit rules defined by the state. Here, blockchain becomes a technology for the state to cut down the actors involved in the process. From pages 12-14, it presents the cost-effective and economic value of 'reducing intermediaries' and decentralization while adding that the "mechanism of decentralization or peer-peer exchange is a spectrum and not a binary concept" and decentralised infrastructure is not opposed to regulation (NITI Aayog, 2020, p. 14).

It is via this caveat that it segues into two characteristics of DLT: "permissioned systems and permission-less systems" (NITI Aayog, 2020, p. 14). Thereafter, a case is made to pay attention to "sectors of governmental intermediation where a state entity is involved just for ledger maintenance or collecting state dues but is not adding value to the transaction" to "assess how government's role can be defined in those sectors" (NITI Aayog, 2020, p. 14). This scalar approach to looking and getting rid of its own organs that do not add value becomes interesting in light of Gupta's (2006) anthropological lens that describes the "state as a multilayered, contradictory, translocal ensemble of institutions, practices, and people in a globalized context [rife with] interbureaucratic conflicts" (2006, p. 16) to underscore the porous, permeable and heterogeneous nature of the state and its institutions which distributes power and creates space for layers of contestations. Moreover, the dismissal of permissionless or public blockchains further points towards the erasure of (pseudo)anonymous transparency that DLT could facilitate.

To sum up, with Bitcoin, blockchain represented a move away from a trust-based system towards transparency, whereas in the NITI Aayog's re-description, blockchain is a move towards a code-based intermediary that allows the state to reorganize itself and re-enter as a trusted intermediary.

Having made this discursive shift from a) transparency to trust to blockchain as adjustably transparent, b) transaction as any exchange/deal (not just financial) between parties and blockchain as a database/record of that exchange, and c) from disintermediation to reduced intermediaries, the discussion paper goes on to describe its blockchain use-case selection framework.

It presents a list of nine items to consider before settling on a blockchain-based solution. Of these nine items, the foremost is the need to reduce intermediaries and the number 5 points the need for shared write access.

We point out number 5 to underscore the discussion paper's emphasis on shared write access as a necessary condition to adopt blockchain-based solutions as opposed to shared read access. Within its framing of reducing intermediaries, it suggests that if multiple actors do not require write-access on the ledger then blockchain is not the right fit. The question of read access is reduced herein to disintermediation, instead of opening up the ledger to the larger public. As a government infrastructure, transparency on blockchain will be privileged access, with the state returning as a centralized 'trusted party' that decides who gets the write and read access; it does not concern the possibility for a citizen to simply browse the blockchain or have access to the information without going through a 'trusted intermediary', for example, the way one does on Etherscan to browse the transaction data of Ethereum.

The rest of the discussion paper narrates various trials and experiments that NITI Aayog has been pursuing with several other actors to determine the suitability of blockchain deployment in various state-led use cases. It ends with "Regulatory and policy considerations for evolving a vibrant blockchain ecosystem" such as "IndiaChain", establish "India as blockchain hub", "Procurement process for government agencies to adopt blockchain solutions", "Pegged stable coin for Indian Rupee" and the regulatory mechanisms around cryptocurrency and initial coin offerings (ICO) (NITI Aayog, 2020, p. 52).

The absence of NITI Ayog's promised second paper on cryptocurrencies limits our current foresight into the possible configuration of the Indian state and its particular reconfiguration concerning cryptocurrency and blockchain. However, through a close reading of the first part, we have identified the discursive shift from transparency to trust to adjustable transparent that is at the heart of re-describing blockchain as governance infrastructure. Rather than decentralize, this discursive shift will have centralizing effects on the way the technology comes to be designed and deployed with the state returning as a gatekeeper.

If the transparency feature of blockchain was retained as it appears in the 2008 Bitcoin paper, one could have, for example,

imagined a range of state processes, that already fall under the ambit of India's Right to Information Act, to be auditable by the citizens. Activist notions of auditing flourished since the 1950s (Vecchione *et al.*, 2021) which have had concrete effects on social movements and the state with provisions such as the Right to Information in India.

The origins of the Right to Information movement deserve a brief mention here to point out why this erasure is a significant one in a genealogical sense. The Right to Information as citizen-centric auditing of state processes began in a small village in Rajasthan with an initial demand for the panchayat (village governing body) to write their financial accounts on the blackboard outside the office (Roy and MKSS Collective, 2018) which could be read by anyone. This was a vision of a citizen-centric auditable ledger long before DLT.

However, such an imagination is foreclosed in the above discursive shift from transparency to trust to adjustably transparent which clears the space for the state to return as a centralized, trusted intermediary. We acknowledge that thorny issues concerning privacy and blockchain require concerted efforts (Halpin and Piekarska, 2017)and one could argue that opening all ledgers to the public raises privacy-related questions. The report explicitly warns against storing private/proprietary information on blockchain and suggests that it is best suited for transaction records. While transaction records need not necessarily require disclosure to all citizens, the foreclosure of any possibility to use blockchain to open up state records as already accessible information cannot be ignored.

*MeitY's Report*

As noted previously, NITI Aayog is an advisory body to the Union government but has no constitutional power. On the other hand, the Ministry of Electronics and Information Technology (MeitY) is a ministry of the Government of India formed in 2016 when it was separated from the Ministry of Communications and Information Technology. In effect, NITI Aayog is an advisory body whose documents wield advisory power, but a Ministry is a site of executive powers, whose reports have sovereign legitimacy. This power shift has effects on the nature of the reports and their redescription of blockchain as well. While NITI Aayog makes narrative manoeuvres to delineate and redescribe the object under contestation, the ministry moves towards its stabilisation and legitimisation.

*National Strategy on Blockchain* was published in January 2021, a year after the NITI Aayog discussion paper, by the MeitY to "make India as one of the leading countries [to] harnessing" (Ministry of Electronics and Information Technology, Government of India, 2021, p. 2) technology's benefits by focusing on both technological and administrative aspects of blockchain. While overlapping in scope with the NITI Aayog's report, MeitY's report is more straightforward in its tone than the discussion paper. It begins with an emphasis on developing "trusted digital platforms through shared Blockchain infrastructure… and facilitating state of the art, transparent, secure, and trusted digital service delivery to citizens and businesses" (Ministry of Electronics and Information Technology, Government of India, 2021, p. 1). It is divided into 12 chapters over 43 pages to provide insight into strategies and recommendations for creating a trusted digital platform using Blockchain and MeitY's role therein. The document, written in a more matter-of-fact tone, does not deploy the rhetorical moves as explicitly as the NITI Ayog's discussion paper.

It begins by stating in a matter-of-fact manner that "Blockchain is an innovative distributed ledger technology… introduced in the design and development of cryptocurrency, Bitcoin in 2009 by Satoshi Nakamoto…is an amalgamation of various innovations, with a clear business value…enables a shared ledger among the various parties involved in business transactions…acts as the single source of truth…and eliminates the need for a central entity to validate the transactions…can be used in both permissioned and permissionless models" (Ministry of Electronics and Information Technology, Government of India, 2021, p. 2). Like the discussion paper, it presents blockchain as a technology that enables a layer of trust and goes into its technical details at the outset. It reiterates the discursive shift noted in the discussion paper, where under the section on "value addition of the technology in e-governance domain", it states that "…Blockchain follows secure by design paradigm which makes it a unique system that makes business transactions transparent and trusted in a consortium environment" (Ministry of Electronics and Information Technology, Government of India, 2021, p. 7). However, in its analysis of blockchain, it presents a wider use-case for blockchain in a range of domains ranging from cryptocurrencies, trade finance, legal, real estate, healthcare, supply chain, IT cloud storage, sharing economy, Internet of things, insurance, crowd-funding, and banking (Ministry of Electronics and Information Technology, Government of India, 2021, p. 14) and goes on to identify international examples of blockchain-as-a-service in other regions such as China, Europe,

Estonia, United Arab Emirates, US, Brazil, Chile, Canada, Singapore, Switzerland as well as initiatives in the private sectors led by Samsung, LG and Amazon before it goes on to describe the national initiatives.

This allows it to present a wider use case to describe blockchain as a service with economic and social value and remain open to "find new innovative use-cases based on blockchain" (Ministry of Electronics and Information Technology, Government of India, 2021, p. 26). However, when read side-by-side with the NITI Aayog's discussion paper, one can identify the foreclosure underpinning this report. The new use cases are situated in the discursive shift that the NITI Aayog's discussion paper has already enacted. The matter-of-factness is, first and foremost, grounded in this discursive shift from transparency to trust to adjustably transparent.

Where NITI Aayog's report focuses on introducing and explaining the use cases of blockchain for the state and framing the potential of blockchain as an infrastructural intervention for disintermediation, MeitY's report is more concrete in its visions and suggestions. As part of the technological aspects of blockchain, it begins at the outset by recommending a National Blockchain Infrastructure for hosting a regulatory sandbox that can be used for building and deploying Blockchain applications. In addition to the discursive shift, this concrete and matter-of-fact tone is made possible via two other stabilising gestures.

The report begins by crediting Bitcoin to Satoshi Nakamoto, which is used in the report as a proper name, as an individual inventor of the technology. This matter-of-factness rests on a discourse of the technological prowess of great individuals in the narratives of technoscientific progression. Noting the presence of technology 'Czars' as an important feature in the postcolonial Indian technoscience narratives, Abraham (2017) argued that this practice "derives from the distinct notion of authorship…as well as the teleological imbrication of higher stages of modernization with technological achievement".

Unlike the discussion paper, which acknowledged the contingencies concerning blockchain technology and the need to redescribe it before it can be used as a governance infrastructure, the matter-of-fact attitude of MeitY's report erases the pseudonymity surrounding the name Satoshi Nakamoto (which can refer to one person or a group of persons) and the contingency around blockchain. The technology is stabilized by assigning it a distinct author, a proper name, and second by narrating it as part of the inevitable higher stages of modernization to come— that is, "an amalgamation of various innovations, with a clear business value" (Ministry of Electronics and Information Technology, Government of India, 2021, p. 1).

Moreover, the report also emphasizes suggestions and feedback from public consultations that map the interest of various actors in the technology and the desire among different actors to steer it in what they consider to be morally and societally right directions, covering aspects such as research and development, security and privacy, standards, legal and regulatory compliance, financial inclusion, education, awareness and skill building etc. A (techno)scientific fact or object is stabilized by enrolling new allies and assembling an authoritative interest (Latour, 1987) of different actors in blockchain technology as potential government infrastructure. Opening up blockchain technology to these other external aspects ranging from legal compliance to skills and competence enacts a stabilizing gesture where blockchain emerges as a code-based trusted intermediary in service of the state.

**Conclusion**

A critical look at how the state redescribes emerging technologies such as blockchain helps foreground the contested nature of technologies. The discussion paper by NITI Aayog enacts the discursive shift from transparency to trust to adjustably transparent that allows the state to re-enter as a trusted intermediary managing and regulating a permissioned blockchain model which is designated decentral but has centralizing effects with the state removing its heterogeneous organs that presumably do not add value and deciding who has the read or write access. MeitY's report furthers this discursive shift and stabilizes blockchain technology as a potential governance infrastructure through its matter-of-fact presentation of blockchain "as an amalgamation of various innovations" and how it has enrolled interests of a range of actors to develop it as a techno-managerial solution that can be leveraged to deliver state's necessary functions. The two documents examined in this article illustrate that the redescription of blockchain away from transparency to trust to adjustably transparent is not an innocent manoeuvre but at the heart of re-describing blockchain technology as a key government infrastructure.

The question vis-à-vis blockchain is not whether it can become a well-regulated general-purpose technology underpinning state or market infrastructure but rather what does it become as processes of re-description continue. These de/recentralization tendencies are not unique to blockchain but form the core of distributed technologies which researchers should watch out for, and critically interrogate. Future research could consider approaching the 'discursive infrastructuring' (see Bennani-Taylor (2024) and Rao (2023)) concerning blockchain and other aspects of scripting a technological artefact by investigating how various actors, institutions and imaginaries are being mobilized and generated around blockchain in real-world scenarios. Ethnographic and larger studies of the emerging blockchain ecosystem in India are exigent as the technology roll-out accelerates and affects different domains of the state and the everyday life of the citizenry.

**About the authors:**

Debarun Sarkar, PhD, is an independent researcher.

Cheshta Arora, PhD, is a senior researcher at the Western Norway Research Institute, Sogndal.

**Acknowledgements**

The first author received no financial funding for his contributions to this paper. The second author received partial financial support from the Western Norway Research Institute in Sogndal, Norway.